# *Key factors and network model for location-based cultural mobile game design*

**Ruo-Yu Li, Chang-Hwa Wang**

*Ruo-Yu Li is a Master graduated from the Department of Graphic Arts and Communications at the National Taiwan Normal University. Her research focuses on game-based learning, mobile learning, cultural recognition and architectures for educational technology system. Chang-Hwa Wang is a Professor in the Department of Graphic Arts and Communications at the National Taiwan Normal University. His research interests include context-aware games and learner preceptions in mixed-reality learning environmenst. Address for Correspondence: Chang-Hwa Wang, Department of Graphic Arts and Communications, National Taiwan Normal University, 162, Section 1, Heping E. Rd., Taipei City 106, Taiwan. Email: wangc@ntnu.edu.tw*

## Abstract

The use of smart devices as media for digital learning constitutes a new-generation digital learning paradigm. Therefore, context-aware game-based learning has attracted considerable attention. Location-based games have not only positive effects on learning but also pronounced effects on culture and history. Accordingly, focusing on railway cultural heritages, we attempted to assess interdependent relationships between key factors crucial for the design of a location-based mobile game for cultural heritages. We adopted the analytic network process (ANP) for our assessment. We initially performed a literature review to generalize relevant criteria and elements and developed a questionnaire based on the fuzzy delphi method (FDM); thus, key factors, namely 3 criteria and 15 elements, were selected. We also applied an online ANP-based questionnaire; on the basis of the experts' opinions, we established a network model and determined the priority order of the key factors. The results revealed that experts considered "culture learning" to be of the highest importance, with the most important three elements being "prior knowledge", "challenge levels," and "cultural narrative." In addition, culture learning exhibited a strong interaction with content design. In each criterion, one element had a considerable influence on the remaining elements, as determined by an analysis of matrices.

## Introduction

Through the context-aware applications of smart devices, mobile learning can enable learners to interact with external environments and capture the corresponding learning materials. Liu and Hwang (2010) indicated that ubiquitous learning has become a new paradigm of digital learning and has influenced the general trend of the future learning field. Studies have confirmed that context-aware game-based learning strategies, such as those involving the use of instructional pervasive games (IPGs), have positive effects on education quality. Specifically, studies have revealed that IPGs could improve learning outcomes, motivation, and achievements (e.g., Huizenga, Admiraal, Akkerman, & Dam, 2009; Liu & Chu, 2010; Su & Cheng, 2015).

## Practitioner Notes

What is already known about this topic

- Context-aware game-based learning, such as IPGs, improves quality of learning.
- Location-based games benefit historical and cultural learning and help learners achieve better learning outcomes, learning potential, and learning interest.
- Most studies on location-based, cultural exploration games mainly emphasize their design and development of the game scheme. However, the key factors that make a success of the game are overlooked.

What this paper adds

- We identified key factors (3 criteria; 15 elements) and developed an ANP-based network model for location-based cultural mobile game design.
- The priority order of key factors for designing a location-based railway cultural game is derived.
- The strengths, relationships, and influences among identified key factors were analyzed.

Implications for practice and/or policy

- The results of this study provide validated guidelines for further development of effective location-based cultural exploration games.
- Our study findings enable designers to prioritize or demote reference factors for construction, according to relative strengths and weaknesses of design factors.

In contrast to general pervasive games or context-aware games, cultural heritage games target certain areas with abundant cultural information; this can motivate players to perceive cultural heritage elements from a different perspective, enabling the construction of personal narratives while creating a personal exposition (Coenen, Mostmans, & Naessens, 2013). Gentes et al. (2010) proposed that IPGs should emphasize the narrative content of urban culture and that pervasive games of urban cultural heritages should be developed from a humanistic perspective.

Scholars have discovered that location-based learning games focusing on historical heritages could considerably improve learning outcomes, learning potential, or learning interest (e.g., Avouris & Yiannoutsou, 2012; Ebling & Cáceres, 2010; Rubino, Barberis, Xhembulla, & Malnati, 2015). In addition, with active advancements in mobile devices, various interactive systems have been developed at cultural heritage sites to assist visitors in understanding cultural heritages (Economou & Meintani, 2011). Ihamäki (2014) suggested that future research can consider the combination of location-based games and AR technology.

However, few studies have focused on key factors associated with the design of location-based cultural mobile games and the interdependent relationships between such factors. Specifically, studies have mostly emphasized designing pervasive games without focusing on key factors for the design of such games; instead, researchers in such studies have only summarized factors through literature reviews and personal experiences (e.g., Gustafsson, Bichard, & Combetto, 2006; Jegers, 2009; Walther, 2005). Moreover, in location-based pervasive games, features such as the Global System for Mobile Communications (GSM) and location-based services may be generalized or ignored. Accordingly, one of the objectives of the current study was to explore key factors (namely criteria and elements) for the design of location-based games for cultural recognition.

For designing IPGs, studies have incorporated educational theoretical frameworks into diverse design models; however, design factors for such games have been referenced from earlier studies (Gentes, Guyot-Mbodji, & Demeure, 2010). Factors associated with cultural recognition are also affected by reliability and validity concerns (Chen & Shih, 2012). Therefore, another objective of the current study was to construct a new network model comprising data collected from experts in three different fields. As mentioned, most scholars have not probed the interdependent relationship or priority order of key factors for game design (Chen, Guo, & Shih, 2012).

From the perspective of game developers, making trade-offs between elements or criteria during game design might be impractical. To address this concern, Saaty (1996) proposed the analytic network process (ANP), which can be used to solve the problems of determining the interdependence between design elements or factors; through the ANP, scholars can design questionnaires to collect the opinions of experts from different fields about key design factors, evaluate the questionnaire data to determine the priority order of the key factors, and determine interdependent relationships between the factors (Saaty, 2016). Accordingly, the current study applied the ANP to determine the priority order of key factors for the design of location-based cultural mobile games; railway cultural heritages were considered in this study.

## Literature review

In the first phase of our research project, as reported in Li and Wang (2018), we applied the FDM to summarize key factors for the design of a location-based cultural mobile game; the factors comprised 4 criteria—namely "culture learning," "context requirements," "system structure," and "content design"—and 20 elements.

### *Culture learning*

Culture learning refers to the inclusion of essential points that be learned in cultural pedagogy. From the perspective of educators, Laine, Sedano, Joy, and Sutinen (2010) proposed developers should collect a sufficient amount of user data in advance; the use of an integrated and reliable evaluation method is also crucial for IPGs. In their discussion of gaming methods, Gentes et al. (2010) indicated that team exploration is more favorable than individual exploration for instructional location-based games. Furthermore, researchers have considered that the cultural content should be designed with input from people who have local cultural knowledge (Chen et al., 2012; Gentes et al., 2010). Cultural narratives also constitute a crucial element of game development; providing such narratives can help players develop an advanced organizer of cultural heritages (Chen & Shih, 2012).

### *Context requirements*

Context requirements refer to features that be exhibited by pervasive computing devices. Context requirements comprise the following four elements that constitute the four axes of pervasive gaming proposed by Walther (2005): "distribution," "mobility," "persistence," and "transmediality." Distribution refers to the ability to distribute gaming information extensively through networks. Additionally, Gustafsson et al. (2006) proposed a concept called "scalability," similar with the meaning of distribution, which was among the design elements in an experimental location-based game for car tours. Mobility refers to the flexibility of pervasive games (Chen & Shih, 2012). Moreover, persistence refers to constant availability of pervasive games (Chen et al., 2012). Transmediality refers to the transfer process between media. Chen and Shih (2012) suggested that new media can connect virtual community networks in multiple manners.

*System structure*

System structure refers to the overall system configuration and involves physical and virtual designs. Such designs comprise several elements, including game mechanics. Game mechanics refer to processes that can enable game engines to monitor and correct virtual and realistic links (Walther, 2005); Walther also indicated that the system structure should comprise adequate game entities. Because a game is a rule-based formal system with a variable and quantifiable outcome (Jasper, Copier & Raessens, 2003), game rules should be developed such that they render cheating difficult and uphold equal opportunities for competition (Laine et al., 2010).

On the basis of the study conducted by Sweetser and Wyeth (2005), Jegers (2009) developed the Pervasive GameFlow Model, which emphasizes that pervasive games should support players as they switch their concentration between tasks of gameplay and surrounding factors. Neustaedter, Tang, and Judge (2013) conducted a study on game-based activities in geocaching and reported that games should enable players to develop elaborate and lightweight creations. Ihamäki (2014) also suggested that game developers incorporate components of mixed reality, such as augmented reality technology, into games in order to enhance players' gaming experience.

*Content design*

Content design refers to the content factors of pervasive games. Jasper (2003) considered that games should encourage players to invest effort to influence game outcomes. Regarding gaming experience, Jacob and Coelho (2011) suggested that location-based games should provide a balanced gaming experience or customization. Regarding game background, location-based pervasive games should have a sequential storytelling system (Gustafsson et al., 2006). In addition, game immersion is a key factor in game design and has received considerable attention. Chen et al. (2012) and Sweetser and Wyeth (2005) have considered that players should be able to effortlessly immerse themselves in the concept of gaming experiences.

According to the GameFlow Model proposed by Sweetser and Wyeth (2005), games should provide clear goals for players at the right time. On the other hand, Jegers (2009) proposed that pervasive games should prompt players to shape and express their own goals; players should also receive feedback on their progress toward their goals. Jegers also considered that pervasive games should enable game-oriented and meaningful interactions within the gaming system. Ihamäki (2014) indicated that an adventure mobile game should support multiplayer competition and information sharing on social media platforms.

## Methods

*Proposed model*

This study involved two phases, namely phase 1 (Li & Wang, 2018) and phase 2. To streamline the literature review in phase 1, we referred to the research model proposed by Wei and Chang (2008) and adopted the fuzzy Delphi method (FDM) to design our questionnaire. After reviewing questionnaire responses, we determined that the experts established 3 criteria and 15 elements, as presented in Table 1.

*Table 1: Key factors for design of location-based mobile game for cultural heritages*

| Criterion | | Element | | Definition | Reference |
|---|---|---|---|---|---|
| **C1** | Culture learning | e11 | Prior knowledge | Understand player backgrounds to measure the learning strategies. | Laine et al., 2010 |
| | | e12 | Team exploration | Integrate team power in problem-solving. | Chen & Shih, 2012; Gentes et al., 2010 |
| | | e13 | Outcomes evaluation | Evaluate learning outcomes with a trusted analytical approach. | Laine et al., 2010 |
| | | e14 | Cultural narrative | Game contents are provided by the local culture. | Chen & Shih, 2012; Gentes et al., 2010 |
| **C2** | System structure | e21 | Game mechanics | Monitor dynamic and modify virtual links. | Neustaedter et al. 2013; Walther, 2005 |
| | | e22 | Game entities | The abstract object that can be moved and drawn over a game map. | Gustafsson et al., 2006; Walther（2005） |
| | | e23 | Mixed reality | Achieve seamless integration of virtual and real world. | Ihamäki, 2014; Jegers, 2009 |
| | | e24 | Game rules | Formulate fair rules such as playing time. | Jasper et al., 2003; Jegers, 2009; Walther, 2005 |
| | | e25 | Control skills | A platform that players can easily control and get started with. | Jegers, 2009; Neustaedter et al., 2013; Sweetser & Wyeth, 2005 |
| **C3** | Content design | e31 | Challenge levels | Consider factors to design the appropriate level of challenge. | Jacob & Coelho, 2011; Jasper et al., 2003 |
| | | e32 | Story arrangement | Sequential story arrangements and design of climactic or conflicting plots. | Gustafsson et al., 2006; Ihamäki, 2014 |
| | | e33 | Game immersion | Allow players to focus on the surroundings and feel immersed. | Chen et al., 2012; Jegers, 2009; Sweetser & Wyeth, 2005 |
| | | e34 | Clear goals | The goals need to be clear and presented in advance. | Jegers, 2009; Sweetser & Wyeth, 2005 |
| | | e35 | Game feedback | Players receive timely and relevant feedback as they move toward the goal. | Jegers, 2009; Sweetser & Wyeth, 2005 |
| | | e36 | Competition and interaction | Triggering contexts encourage players to interact or compete with each other. | Ihamäki, 2014; Jegers, 2009; Sweetser & Wyeth, 2005 |

*Note.* modified from Li and Wang (2018).

Subsequently, on the basis of the result in phase 1, we developed an ANP model based on that developed by Chen, Shih, Shyur, and Wu (2012), as presented in Figure 1, where two-way arrows represent external relationships between the criteria and curved arrows represent internal relationships between elements belonging to the same criterion. We applied the ANP model to design an online questionnaire with a total of 195 questions (see Appendix A), and distributed copies of this questionnaire to seven experts covered both production and education domains in phase 2. Finally, a series of pairwise comparison matrices were constructed for each criterion and element, including the goal of current study, namely “design a railway culture location-based game;” thus, supermatrices were established and analyzed to obtain the weight of each criterion and element.

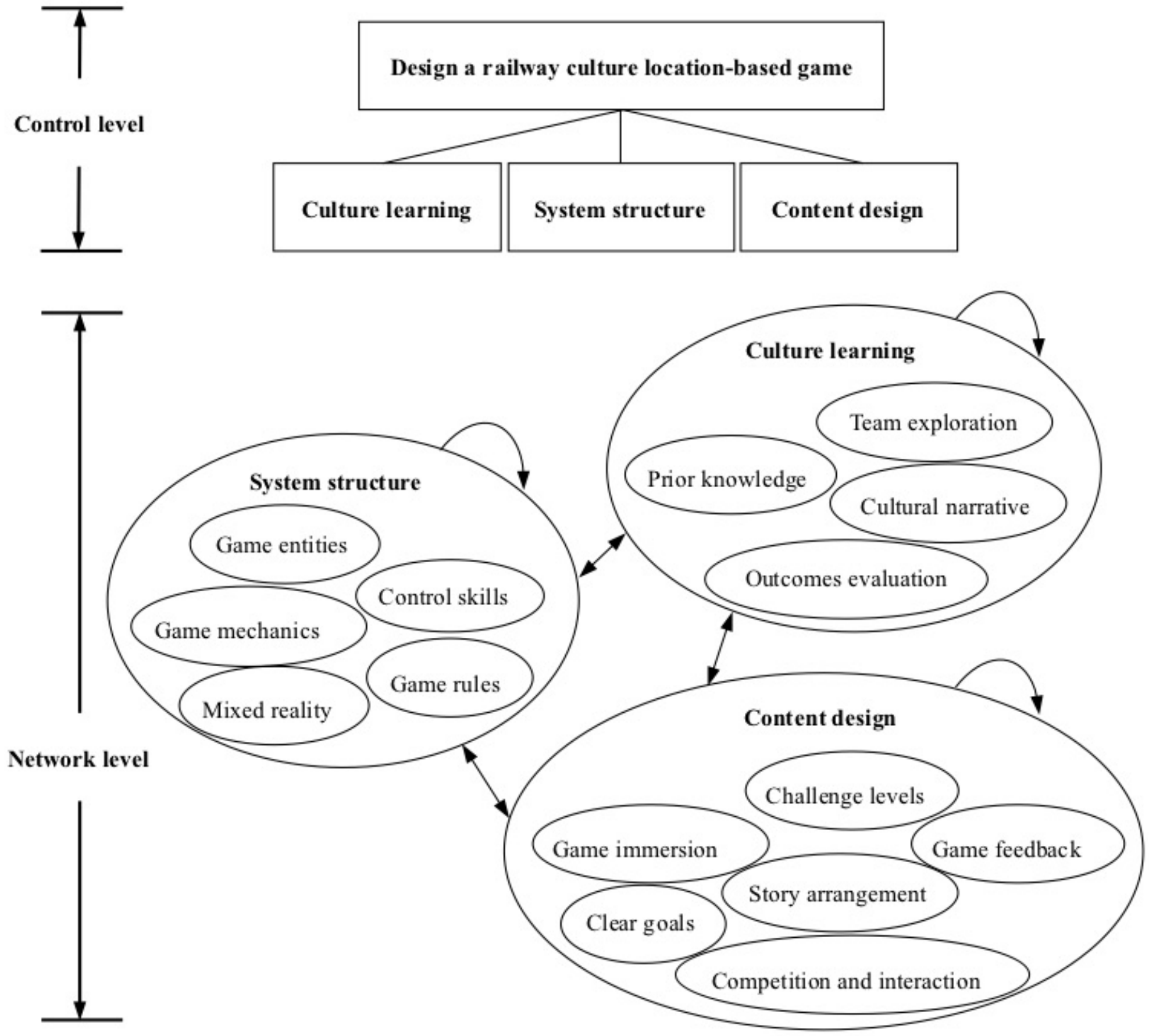

*Figure 1: ANP model for the design of a location-based game for railway cultural heritages*

*Data analysis methods*

Step 1. Constructing pairwise comparison matrices and calculating a priority vector and largest eigenvalue of each factor

After collecting questionnaire responses, we evaluated the relative importance of the criteria and elements on the basis of the 9-point scale proposed by Saaty, Figueira, Greco, and Ehrogott (2005). Pairwise comparison matrices of the criteria and elements can be expressed as follows:

$$A = \begin{array}{c} \\ A_1 \\ A_2 \\ \vdots \\ A_n \end{array} \begin{array}{c} \begin{array}{cccc} A_1 & A_2 & \cdots & A_n \end{array} \\ \begin{bmatrix} w_1/w_1 & w_1/w_2 & \cdots & w_1/w_n \\ w_2/w_1 & w_2/w_2 & \cdots & w_2/w_n \\ \vdots & \vdots & \cdots & \vdots \\ w_n/w_1 & w_n/w_2 & \cdots & w_n/w_n \end{bmatrix} \end{array} \tag{1}$$

$$= \left[a_{ij}\right] = \begin{bmatrix} 1 & a_{12} & \cdots & a_{1n} \\ a_{21} & 1 & \cdots & a_{2n} \\ \vdots & \vdots & \cdots & \vdots \\ a_{n1} & a_{n2} & \cdots & 1 \end{bmatrix} \tag{2}$$

$A$ is an $n \times n$ square matrix, where $A_1, A_2, \cdots, A_n$ represent $n$ elements in a criterion; $w$ represents criterion $j$ or element $j$ evaluated using the 9-point scale under the premise of another criterion $i$ or factor $i$. In addition, $a_{ij} = w_i / w_j$. Because the importance value was equal to the value of the element itself, each of the values along the diagonal of the matrix was 1; this diagonal divided the matrix into upper and lower triangles, with the values in the lower triangle being the reciprocal of those in the upper triangle.

Subsequently, we calculated the geometric mean of each column vector in the matrix and then approximated and normalized it, as indicated by the following equation:

$$w_i = \frac{\sqrt[n]{\prod_{j=1}^{n} a_{ij}}}{\sum_{i=1}^{n}\left(\sqrt[n]{\prod_{j=1}^{n} a_{ij}}\right)}, \quad \forall i, j = 1,2,\dots,n \tag{3}$$

Step 2. Testing consistency using the consistency index and consistency ratio

To control the transitivity within an acceptable range during the pairwise comparison, we calculated the consistency index (C.I.) by using the following equation proposed by Saaty (1996):

$$C.I. = \frac{\lambda_{max} - n}{n-1} \tag{4}$$

where $\lambda_{max}$ represents the maximum eigenvalue and $n$ represents the number of elements. If the C.I. is ≤0.1, which is within the allowable error range, a comparison matrix can be considered to achieve consistency. Furthermore, we applied the consistency ratio (C.R.) proposed by Saaty (1996) to test the consistency of each matrix C.I. and the random index of the matrix in same rank. The calculation of C.R. first looked up the Table 2 for the random index (R.I.).

*Table 2: R.I. comparison table*

| *Rank* | 1 | 2 | 3 | 4 | 5 | 6 | 7 | 8 |
|---|---|---|---|---|---|---|---|---|
| $R.I.$ | 0.00 | 0.00 | 0.58 | 0.90 | 1.12 | 1.24 | 1.32 | 1.41 |
| *Rank* | 9 | 10 | 11 | 12 | 13 | 14 | 15 | |
| $R.I.$ | 1.45 | 1.49 | 1.51 | 1.48 | 1.56 | 1.57 | 1.58 | |

Saaty *et al.* (2005)

The C.R. formula is presented as follows:

$$C.R. = \frac{C.I.}{R.I.} \tag{5}$$

A C.R. value of ≤ 0.1 (and a corresponding C.I. value of ≤ 0.1) was considered to indicate satisfactory consistency in this study.

Step 3. Constructing a supermatrix and calculating overall limit weight values

We derived a supermatrix $W$, which is expressed as follows:

$$
W = \begin{array}{cc} & \\ C_1 & \begin{array}{c} e_{11} \\ e_{12} \\ \vdots \\ e_{1n} \end{array} \\ C_2 & \begin{array}{c} e_{21} \\ e_{22} \\ \vdots \\ e_{2n} \end{array} \\ \vdots & \vdots \\ C_k & \begin{array}{c} e_{k1} \\ e_{k2} \\ \vdots \\ e_{kn} \end{array} \end{array}
\begin{array}{c} \begin{array}{ccc} C_1 & C_2 & \cdots & C_k \\ e_{11}\ e_{12} \cdots e_{1n} & e_{21}\ e_{22} \cdots e_{2n} & \cdots & e_{k1}\ e_{k2} \cdots e_{kn} \end{array} \\ \begin{bmatrix} w_{11} & w_{12} & \cdots & w_{1k} \\ w_{21} & w_{22} & \cdots & w_{2k} \\ \vdots & \vdots & & \vdots \\ w_{k1} & w_{k2} & \cdots & w_{kk} \end{bmatrix} \end{array} \quad (6)
$$

where $k$ denotes the number of criteria, presented as $C_1, C_2, \dots, C_k$, with $C_k = \{e_{k1}, e_{k2}, \dots, e_{kn}\}$ and $e$ representing the number of elements $n$ in $C_k$. Each supermatrix $W$ has a submatrix $w$, the unit of which can be derived from the priority vector of a pairwise comparison matrix. Matrix $W^h$ along with its submatrices can be expressed as follows:

$$
W^h \quad = \begin{bmatrix} 0 & 0 & 0 \\ w_{21} & w_{22} & 0 \\ 0 & w_{32} & w_{33} \end{bmatrix} \quad (7)
$$

where $w_{21}$ represents the priority vector of each criterion based on the decision problem, $w_{22}$ represents the priority vector between criteria, $w_{32}$ represents the priority vector of each element based on its criterion, and $w_{33}$ represents the priority vector between elements. After constructing the supermatrix, we multiplied the priority vectors in the unweighted supermatrix by the weights of the pairwise comparison matrices. Thus, we derived a weighted supermatrix $W^a$; in this matrix, the sum of the vectors in each row was 1, and the values of the rows were also random.

Finally, we conducted a power calculation for the weighted supermatrix $W^a$; a limit supermatrix $W^n$ could then be derived as follows:

$$
W^n = \lim_{a \to \infty} W^a \quad (8)
$$

Step 4. Evaluating priorities of key factors and discussing the results

We evaluated the weights of the criteria and elements in the limit supermatrix and then sorted the evaluated weights in the descending order; thus, we could determine the priority order of the key factors. Furthermore, we prioritized partial and overall comparison factors that exhibited the highest importance values; such factors were also regarded as the most important criteria and elements for the construction of location-based games for railway cultural heritages.

## Results

*Consistency tests*

According to the consistency data (see Appendix C), all C.I. and C.R. values were ≤0.1, validating the consistency of the criteria and elements. In addition, calculating the C.R. of the three criteria was not necessary because the R.I. values were all 0.

*Pairwise comparison matrices*

Pairwise comparison matrices were established for three kind, namely "goal," criteria, and elements (see Appendix B). First, goal was established in the control layer to explore the effects of the three criteria: culture learning (denoted by the code C1), system structure (denoted by the code C2), and content design (denoted by the code C3). The weights of matrix established for goal showed C1 had the highest level of importance, followed by C3 and then C2. Second, we assumed that C1, C2, and C3 influenced each other. According to the weights of matrices for the three criteria, C3 had a greater influence on the importance of C1 than did C2. These relationships are presented in Figure 2.

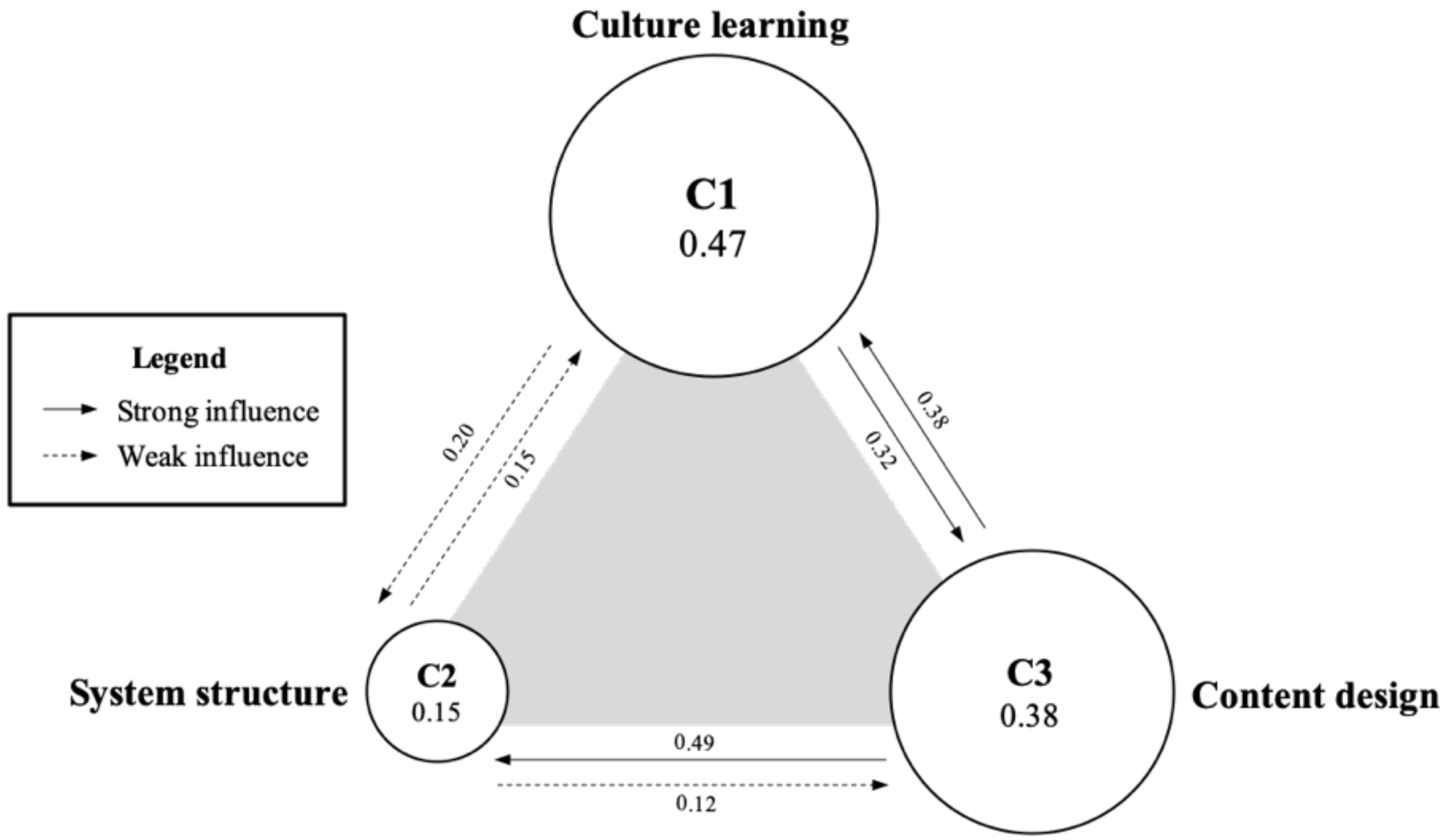

*Figure 2: Strength of the criteria*

As illustrated in the figure, C1 and C3 exhibited a closer interaction.

Third, we established pairwise comparison matrices for the 15 elements on the basis of the premise that one element in a criterion would affect the remaining elements in the same criterion. The effects of all criteria and elements in this study are illustrated in Figure 3.

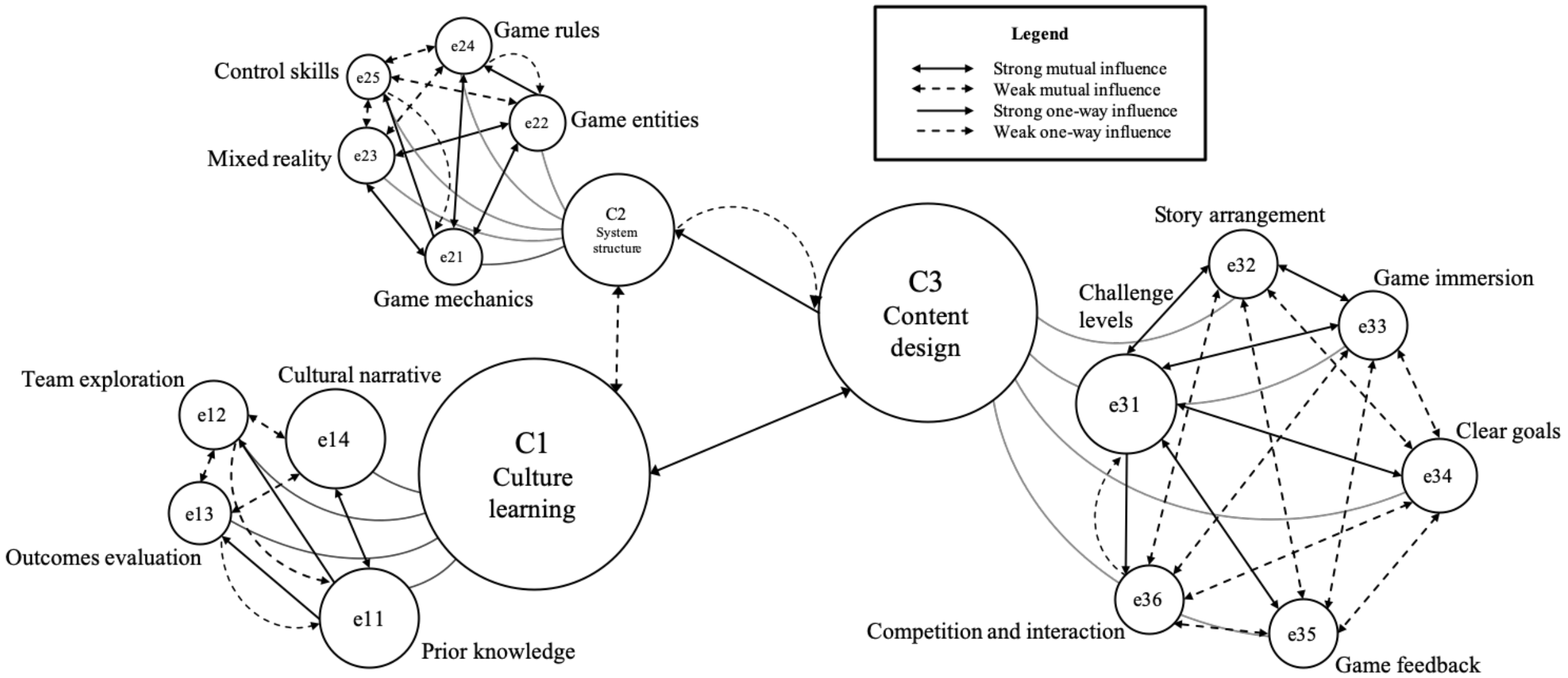

*Figure 3: Strength of the elements*

By observing the relative importance of the elements in C1, we found that e11 had the strongest influence on e12–e14; we also observed a bidirectional relationship between e14 and e11. Furthermore, we observed the relative importance of the elements in C2 and determined that e21 had the strongest influence on the remaining elements in C2. In addition, e22 exhibited the second strongest influence on the remaining elements in C2. By observing the relative importance of the elements in C3, we noted similar results to those observed for the elements in C1 and C2. Specifically, we noted that e31 had the strongest influence on the remaining elements in C3; on the average, the remaining elements, also had a strong influence on e31, except for e36.

In summary, this study determined the degrees of importance of each criterion and element in the study framework, in contrast to previous studies that have rarely analyzed the degree of importance of factors influencing an outcome of interest. Our study findings and procedures can provide clearer information regarding the relative strengths and weaknesses of design factors in order to enable designers to prioritize or demote reference factors for construction.

*Priority order of key factors*

The values of a limit supermatrix (see Appendix D) were defined as the weight values of the overall criterion and element, which served as the basis for the ranking of key factors in this study. We normalized the weight of all key factors and ranked the overall weights in descending order, as presented in Table 3.

*Table 3: Overall priority order of key factors*

| Criterion | | Element | | Overall weight | Overall order |
|---|---|---|---|---|---|
| **C1** | Culture learning | e11 | Prior knowledge | 0.11865 | 1 |
| **C3** | Content design | e31 | Challenge levels | 0.11059 | 2 |
| **C1** | Culture learning | e14 | Cultural narrative | 0.10030 | 3 |
| **C3** | Content design | e34 | Clear goals | 0.08396 | 4 |
| **C1** | Culture learning | e12 | Team exploration | 0.07767 | 5 |
| **C3** | Content design | e32 | Story arrangement | 0.07765 | 6 |
| | | e33 | Game immersion | 0.07105 | 7 |
| | | e35 | Game feedback | 0.07021 | 8 |
| | | e36 | Competition and interaction | 0.06992 | 9 |
| **C1** | Culture learning | e13 | Outcomes evaluation | 0.05769 | 10 |
| **C2** | System structure | e21 | Game mechanics | 0.03859 | 11 |
| | | e22 | Game entities | 0.03467 | 12 |
| | | e23 | Mixed reality | 0.03166 | 13 |
| | | e24 | Game rules | 0.03032 | 14 |
| | | e25 | Control skills | 0.02708 | 15 |

## Discussion

*Key factors for location-based mobile game for cultural heritages*

We conducted this study to determine key factors for location-based cultural mobile games. The study involved two phases. In phase 1, we administered an FDM-based questionnaire to experts for evaluating the relative importance of 4 criteria and 20 elements derived from the literature. Of these criteria and elements, the experts retained 3 criteria and 15 elements; these criteria and elements were considered the key factors in this study. They deleted several criteria and elements, including the criterion context requirements and the following four elements: distribution, mobility, persistence, and transmediality; furthermore, one element, namely collaborative contents, belonging to the criterion culture learning was deleted.

The reasons for the deletions in this study are provided as follows. First, Walther (2005) proposed the four axes of pervasive gaming, and these axes correspond to the four elements that constitute the criterion context requirements; since the execution of this study, considerable technological changes have been achieved. Specifically, in 2005 (i.e., the period of the study conducted by Walther), the iPhone had yet to appear; therefore, smart devices had not reached a mature development stage. In addition, wireless network technology did not have the capacity to support considerably large multimedia data. Accordingly, some of the criteria and elements might have been deleted from this study because of their irrelevance to current smart devices and technologies.

Second, several studies have cited the concept of the four axes of pervasive gaming. However, some studies have modified this concept. For example, Chen, Shih, and Ma (2014), who focused on factors for designing IPGs for cultural recognition, modified this concept on the basis of a literature review, retaining only mobility as one of the design elements in their study. Furthermore, Valente, Feijó, and Leite (2017) described the characteristics of pervasive mobile games; they determined that mobility, persistence, and transmediality are necessary characteristics for confirming pervasive mobile games. As indicated by the aforementioned studies, scholars have modified or partially omitted the four axes of pervasive gaming; this also supports our preceding suggestion that the concept proposed by Walther (2005) is too outdated for current location-based games.

Third, we found that the study by Chen et al. (2014) included fewer factors of game design compared with the current study. Some of the factors used by Chen et al. and the current study might have different names but might overlap in terms of definition; for example, mobility could be regarded as one of the elements belonging to system structure. Accordingly, context requirements and the corresponding four elements might be replaced by other key factors, such as game mechanics and mixed reality, thus explaining the deletion of some of the elements and criteria.

Finally, according to Valente et al. (2017), the four axes of pervasive gaming are suitable for studies reviewing or identifying design factors for the same type of pervasive game. We consider that the concept of the four axes proposed by Walther (2005) might be feasible for the same types of games because each element in Valente's study is fully explained and supplemented by multiple questions, thus preventing misunderstandings or misinterpretations in terms of the name or context of the elements. Accordingly, another reason for the deletion of some of the criteria and elements is that their interpretations were relatively simple, which may cause the experts to misunderstand or fail to determine the precise meanings of the criteria and elements.

*Network structure for and interrelationships between factors for location-based cultural games*

In this study, the participating experts considered that all criteria and elements were interrelated, thus forming a network structure suitable for ANP analysis (see Figure 1). This network model is consistent with models constructed in previous studies. For example, Laine et al. (2010) proposed a technology integration–based model including the factors of instruction, context, and design for IPGs; they also indicated that these factors were interrelated from the perspective of technology integration, and they demonstrated the strengths of such interrelationships in their model. In addition, Chen and Shih (2012) developed a new metamodel emphasizing that context, instruction, and design factors overlapped with each other and were not mutually exclusive. According to these studies, the criteria and elements in the present study were determined to be mutually related and dependent, consistent with the experts' evaluation results.

*Priority order of key factors for development of location-based game for railway cultural heritages*

In general, the priority order of key factors or programs for the design of a product can be determined through a decision-making approach that involves prioritizing the design factors or programs. Through such an approach, the order of the programs or factors for product design can be determined. A similar approach was applied for the prioritization of the factors associated with the design of a location-based game for railway cultural heritages in this study. Specifically, the experts considered that among the three criteria, culture learning was of the highest importance, followed by content design and then system structure.

Among the elements belonging to the criterion culture learning, the experts considered that prior knowledge was of the highest importance, whereas outcome evaluation was of the lowest importance. Additionally, among the elements belonging to the criterion content design, challenge level was of the highest importance, whereas competition and interaction were of the lowest importance. Finally, among the elements belonging to the criterion system structure, game mechanics was of the highest importance, whereas control skills was of the lowest importance. We also observed the gaps in the weight values of the elements.

Our observations revealed that the gaps for the four elements under the criterion culture learning were evenly distributed; by contrast, the gaps observed for the elements under system structure were relatively small, signifying that the elements exhibited similar degrees of importance. In content design, the gaps between challenge levels and the other five elements were relatively large. The preceding results demonstrate that the experts had a clearer priority order for the elements in culture learning. However, they considered that the elements belonging to system structure were not too different with respect to importance. Moreover, according to the experts, challenge level was possibly the most representative element in content design, but the remaining elements were determined to have similar or fuzzy degrees of importance.

Considering the overall priority order of all elements determined by the experts, the top five elements can be ranked as follows in descending order (i.e., from first to fifth): prior knowledge, challenge levels, cultural narrative, clear goals, and team exploration. The bottom five elements (i.e., 11th to the 15th elements) all belonged to criterion system structure: game mechanics, game entities, mixed reality, game rules, and control skills. Notably, the weight values of the top three elements (i.e., prior knowledge, challenge levels, and cultural narrative) were considerably greater than those of the remaining 12 elements. We did not observe significant gap in weight values between the fourth (i.e., clear goals) and ninth (i.e., competition and interaction) elements or between bottom five elements (see Figure 4).

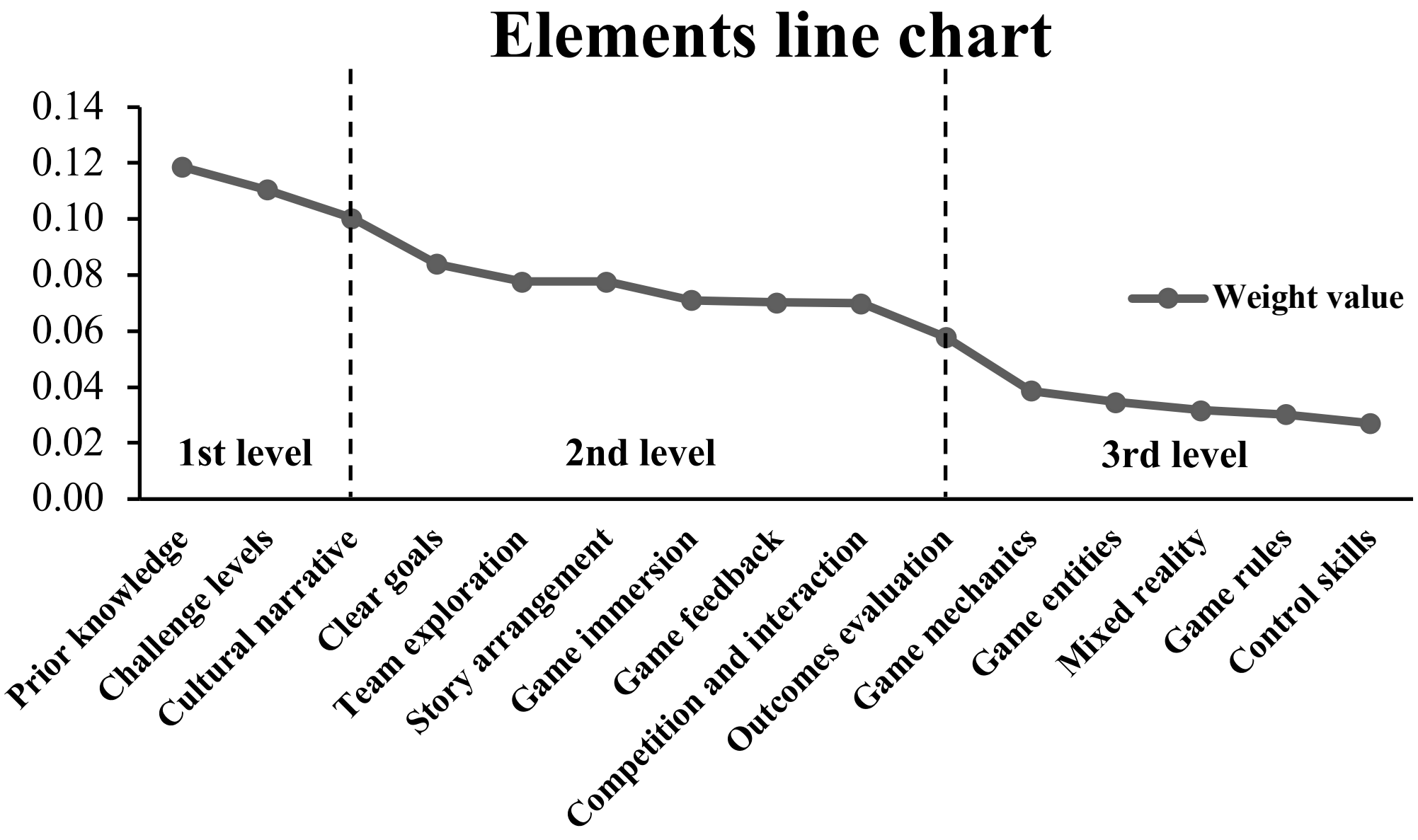

*Figure 4: Line chart of overall weighting values of elements*

This result reveals that the experts assigned the top three elements to the first-level element group, assigned elements with no considerable difference in weight values to the second-level element group, and finally assigned the bottom five elements to the third-level element group.

## Conclusions

This study was performed to identify key factors associated with the design of a location-based mobile game for railway cultural heritages. The study involved two phases, namely phase 1 and phase 2. In phase 1, 3 criteria and 15 elements were selected as the key factors for game design; the following criteria and the corresponding elements were deleted: context requirements (along with the corresponding four elements) and collaborative contents. Three possible reasons for the deletion of these elements and criterion were technological advancements (rendering some of the criteria and elements outdated), changes in meanings of elements, and expert misunderstandings. Furthermore, we constructed network model for the factors, and this model is consistent with those constructed by previous studies.

In phase 2, we determined the priority order of the criteria; specifically, the criteria could be prioritized as follows (in descending order): culture learning, content design, and system structure. We also identified the relative priorities of the elements in each of the criteria. The elements belonging to culture learning had a clear hierarchical order, whereas those belonging to system structure did not have a clear hierarchical order. Moreover, in content design, the most representative element was challenge level; the remaining elements exhibited similar or fuzzy degrees of importance.

This study suggests that the development of location-based mobile games for cultural heritages primarily focus on culture learning and content design. Developers should also prioritize the evaluation of the prior knowledge of each learner to construct an effective learning strategy. Furthermore, challenge levels should be prioritized in game design to ensure that the designed games meet players' different knowledge levels. Developers should also adapt the information of cultural heritages to the narrative content of their designed games, thus tailoring the designed game maps to certain areas.

This study also suggests that games should enable players to realize their goals in advance, using appropriate and immediate feedback to encourage them to achieve their goals. Games should also encourage players to compete with each other during gameplay; this can be achieved by developing a captivating, rich, and moving story that can provide players with an immersive experience while maintaining their awareness of their surroundings. Finally, this study suggests that game system structure should be considered in game design. Specifically, developers should consider game mechanics that can always synchronize a game's environment with other players' environments as well as modifying virtual and realistic links.

## Suggestions for future research

First, this study adopted only quantitative analysis methods to evaluate and present the research results. We suggest that future research use multiple-criterion decision-making methods supplemented with qualitative data (such as interviews) in order to obtain a clearer understanding of collected data. Second, phase 2 of this study included only one expert from industry; this is because companies developing location-based games for cultural heritage learning are rare in Taiwan. Nevertheless, we recommend that future studies use sufficiently large samples and control the number of participants from different backgrounds for the purpose of objectivity.

Furthermore, the results obtained in phase 1 of this study revealed a different priority order of the key factors when compared with the results obtained in phase 2; determining the reason for this discrepancy is warranted. We also found that experts from different fields had slightly different opinions on certain factors. We thus suggest that analyzing the opinions of experts from dissimilar backgrounds is warranted. Finally, according to the result of this study, system structure was of the least importance among the three criteria; each criterion also included one or two elements with relatively weak interdependent relationship.

This thus raises the question as to whether the consideration of numerous factors in the design of location-based cultural games is useful. We hope that researchers can modify or delete ineffective factors by conducting comprehensive literature reviews; doing so can ensure that research obtain factors that are more relevant to the actual requirements of game development.

## Acknowledgements

We would like to give our sincere gratitude to all participating experts in this study. Additionally, this manuscript was edited by Wallace Academic Editing.

## Statement on open data, ethics and conflict of interest

The data used in this study can be accessed by written request to the corresponding author.
We followed the ethical guidelines of corresponding university for data collection. In order to protect the privacy of experts, their names and contact information were not revealed during the research process as well.
There is no conflict of interest resulting from the research work in current study.